\documentclass[11pt, onecolumn]{article}

\usepackage[margin=1in]{geometry} 
\usepackage{amsmath}              
\usepackage{braket}               
\usepackage{graphicx}             
\usepackage{hyperref}             
\usepackage[T1]{fontenc}          
\usepackage{authblk}              
\usepackage{times}                
\usepackage{siunitx}              

\hypersetup{
    colorlinks=true,
    linkcolor=blue,
    filecolor=magenta,      
    urlcolor=cyan,
    citecolor=red,
}

\title{\textbf{Exploring Topological and Localization Phenomena in SSH Chains under Generalized AAH Modulation: A Computational Approach}}

\author[1]{Souvik Ghosh}
\author[2]{Sayak Roy}

\affil[1,2]{Adamas University, Kolkata, West Bengal, India}
\affil[1]{\textit{souvikghosh2012@gmail.com}}
\affil[2]{\textit{sayakroy506@gmail.com}}

\date{June 2025}

\begin{document}

\maketitle

\begin{abstract}
The Su-Schrieffer-Heeger (SSH) model serves as a canonical example of a one-dimensional topological insulator, yet its behavior under more complex, realistic conditions remains a fertile ground for research. This paper presents a comprehensive computational investigation into generalized SSH models, exploring the interplay between topology, quasi-periodic disorder, non-Hermiticity, and time-dependent driving. Using exact diagonalization and specialized numerical solvers, we map the system's phase space through its spectral properties and localization characteristics, quantified by the Inverse Participation Ratio (IPR). We demonstrate that while the standard SSH model exhibits topologically protected edge states, these are destroyed by a localization transition induced by strong Aubry-André-Harper (AAH) modulation. Further, we employ unsupervised machine learning (PCA) to autonomously classify the system's phases, revealing that strong localization can obscure underlying topological signatures. Extending the model beyond Hermiticity, we uncover the non-Hermitian skin effect, a dramatic localization of all bulk states at a boundary. Finally, we apply a periodic Floquet drive to a topologically trivial chain, successfully engineering a Floquet topological insulator characterized by the emergence of anomalous edge states at the boundaries of the quasi-energy zone. These findings collectively provide a multi-faceted view of the rich phenomena hosted in generalized 1D topological systems.
\end{abstract}

\section{Introduction}
\label{sec:introduction}

The discovery of topological phases of matter has revolutionized our understanding of quantum systems, shifting the focus from symmetry-breaking orders to properties protected by the global topology of the bulk electronic wavefunctions \cite{Hasan2010, Qi2011}. A key hallmark of these phases is the emergence of robust, gapless states localized at the system's boundaries, a principle known as the bulk-boundary correspondence. These edge states are immune to local perturbations and defects, making them promising candidates for applications in fault-tolerant quantum computing and spintronics.

Among the pantheon of topological models, the one-dimensional Su-Schrieffer-Heeger (SSH) chain stands out as the simplest and most intuitive example \cite{Su1979}. Originally developed to describe solitons in polyacetylene, it has become a cornerstone of the field, providing a clear illustration of how a topological invariant (the Zak phase or winding number) predicts the existence or absence of zero-energy edge states.

However, real-world physical systems are rarely as clean or static as the pristine SSH model. They are subject to various complex effects, such as intrinsic disorder, coupling to the environment (leading to gain and loss), or external time-dependent fields. The interplay of topology with these effects gives rise to a landscape of rich and often surprising phenomena. This paper embarks on a computational exploration of such "generalized" SSH models, focusing on three key extensions:
\begin{enumerate}
    \item \textbf{Quasi-periodic Disorder:} By incorporating an on-site potential described by the Aubry-André-Harper (AAH) modulation, we study the competition between topological protection and quasi-periodic disorder, which can drive a localization-delocalization transition \cite{Aubry1980}.
    \item \textbf{Non-Hermitian Physics:} By introducing non-reciprocal hopping amplitudes, we move beyond conventional Hermitian quantum mechanics to explore the physics of open systems. This regime is known to host unique phenomena such as the non-Hermitian skin effect, which fundamentally alters the bulk-boundary correspondence \cite{Yao2018}.
    \item \textbf{Floquet Engineering:} By subjecting the system to a periodic, time-dependent drive, we investigate the possibility of creating topological phases that have no static counterpart. This "Floquet engineering" can induce anomalous edge states and dynamically alter the system's topological properties \cite{Oka2009}.
\end{enumerate}

Using a suite of numerical tools, this work systematically investigates the spectral and localization properties of each of these generalized models. We aim to provide a clear, comparative view of how these different physical ingredients reshape the topological landscape of the foundational SSH chain.

The paper is structured as follows: In Section \ref{sec:models}, we define the tight-binding Hamiltonians for the four models under investigation. In Section \ref{sec:methods}, we outline the computational techniques employed. In Section \ref{sec:results}, we present and discuss our numerical results for each model in sequence. Finally, in Section \ref{sec:conclusion}, we summarize our key findings and conclude the paper.

\section{Theoretical Models}
\label{sec:models}

Our investigation is based on the tight-binding approximation. We consider a one-dimensional chain composed of $N$ unit cells, for a total of $2N$ sites, under open boundary conditions (OBC). The Hamiltonian for each model is described below.

\subsection{The Standard SSH Chain}
The standard SSH model describes a dimerized chain with alternating intracell hopping $v$ and intercell hopping $w$. The Hamiltonian is given by:
\begin{equation}
    \hat{H}_{\text{SSH}} = \sum_{j=1}^{N} \left( v \, \hat{c}_{j,A}^\dagger \hat{c}_{j,B} + \text{h.c.} \right) + \sum_{j=1}^{N-1} \left( w \, \hat{c}_{j,B}^\dagger \hat{c}_{j+1,A} + \text{h.c.} \right)
    \label{eq:ssh}
\end{equation}
where $\hat{c}_{j,A}^\dagger$ ($\hat{c}_{j,A}$) creates (annihilates) an electron on the A sublattice of the $j$-th unit cell. The system exhibits a topological phase for $|v| < |w|$, characterized by two zero-energy states localized at the ends of the chain.

\subsection{The SSH-AAH Model}
To study the interplay between topology and quasi-periodic disorder, we introduce an on-site potential with Aubry-André-Harper (AAH) modulation. The full Hamiltonian is $\hat{H}_{\text{SSH-AAH}} = \hat{H}_{\text{SSH}} + \sum_{n=1}^{2N} V_n \, \hat{c}_n^\dagger \hat{c}_n$, where the potential $V_n$ is given by:
\begin{equation}
    V_n = \lambda \cos(2\pi\alpha n + \phi)
    \label{eq:aah_potential}
\end{equation}
Here, $n$ is the site index, $\lambda$ is the modulation strength, $\alpha$ is the inverse golden ratio, and the phase offset $\phi$ is set to zero.

\subsection{The Non-Hermitian SSH Model}
To explore physics beyond Hermitian quantum mechanics, we introduce non-reciprocity in the intracell hopping. The Hamiltonian becomes:
\begin{equation}
    \hat{H}_{\text{NH-SSH}} = \sum_{j=1}^{N} \left( (v+\delta) \, \hat{c}_{j,A}^\dagger \hat{c}_{j,B} + (v-\delta) \, \hat{c}_{j,B}^\dagger \hat{c}_{j,A} \right) + \sum_{j=1}^{N-1} \left( w \, \hat{c}_{j,B}^\dagger \hat{c}_{j+1,A} + \text{h.c.} \right)
    \label{eq:nh-ssh}
\end{equation}
The parameter $\delta$ breaks the Hermiticity ($\hat{H} \neq \hat{H}^\dagger$), leading to a complex energy spectrum and the non-Hermitian skin effect.

\subsection{The Floquet SSH Model}
To investigate time-dependent effects, we apply a periodic drive $\hat{H}(t)$ with period $T = 2\pi/\omega$ to a trivial SSH chain. The evolution is governed by Floquet theory. We implement a five-step discrete drive protocol within each period, defined by $\hat{H}(t) = \hat{H}_{\text{SSH}} + f(t)\hat{H}_{\text{drive}}$, where $\hat{H}_{\text{drive}}$ modifies only the intracell hopping. The time-dependent coefficient $f(t)$ is a piecewise function:
\begin{equation}
    f(t) = \frac{\pi}{2T} \times
    \begin{cases}
        +1 & 0 \le t < T/5 \\
        +1 & T/5 \le t < 2T/5 \\
        -2 & 2T/5 \le t < 3T/5 \\
        +1 & 3T/5 \le t < 4T/5 \\
        +1 & 4T/5 \le t < T
    \end{cases}
\end{equation}
This protocol is known to be capable of inducing a topological phase, characterized by edge states at quasi-energies $E=0$ and $E = \pm \hbar\omega/2$.

\section{Computational Methods}
\label{sec:methods}
The theoretical models were investigated using custom Python scripts leveraging NumPy, SciPy, scikit-learn, and QuTiP. For static Hamiltonians, we performed exact diagonalization of the matrix representation to find eigenvalues and eigenvectors. For the non-Hermitian model, a general-purpose solver was used. To quantify the localization of an eigenstate $\ket{\psi_k} = \sum_{n} \psi_k(n) \ket{n}$, we computed the Inverse Participation Ratio (IPR):
\begin{equation}
    \text{IPR}(\psi_k) = \sum_{n=1}^{2N} |\psi_k(n)|^4
    \label{eq:ipr}
\end{equation}
An IPR value approaching 1 signifies strong localization. For phase classification, we used Principal Component Analysis (PCA) on the probability density vectors $[|\psi(1)|^2, \dots, |\psi(2N)|^2]$ of the eigenstates. For the driven system, we used the Floquet solvers in QuTiP to compute the quasi-energies and Floquet modes.

\section{Results and Discussion}
\label{sec:results}

In this section, we present the numerical results. Unless otherwise specified, a consistent system size of $N=50$ cells (100 sites) was used across the simulations to ensure the comparability of results and mitigate significant finite-size effects.

\subsection{Localization and Topology in the Standard SSH Chain}
\begin{figure}[h!]
    \centering
    \includegraphics[width=0.7\textwidth]{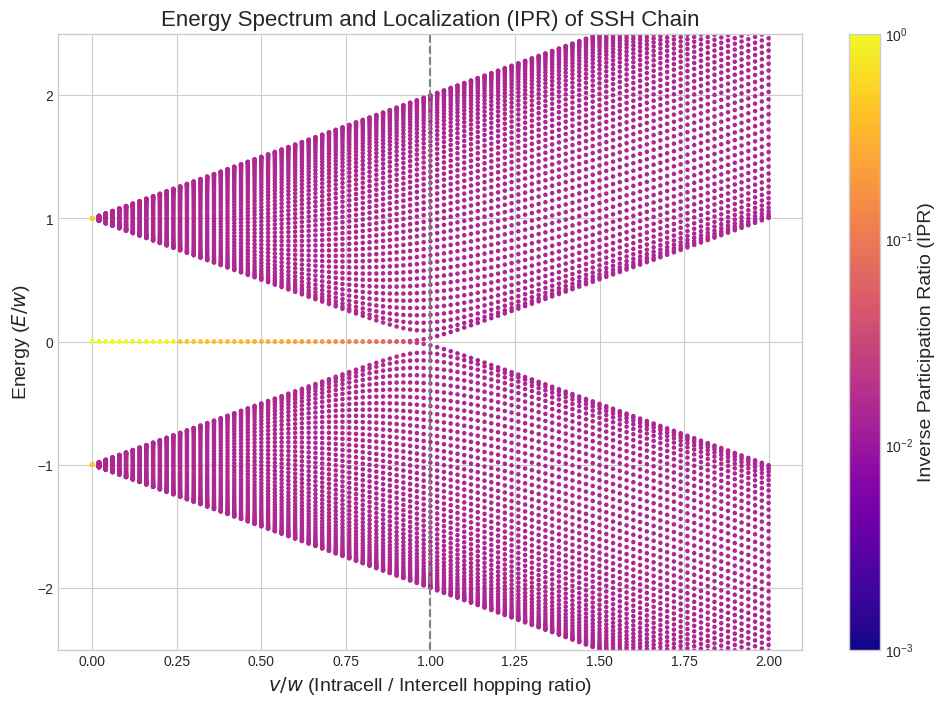}
    \caption{The energy spectrum of the standard SSH chain as a function of the hopping ratio $v/w$, with each eigenstate colored by its IPR. The bright yellow line at $E=0$ in the topological phase ($v/w < 1$) indicates highly localized edge states.}
    \label{fig:ssh_ipr}
\end{figure}

We begin by analyzing the foundational SSH model. Figure~\ref{fig:ssh_ipr} shows the energy spectrum as a function of $v/w$, with each eigenstate colored by its IPR.

The plot clearly illustrates the bulk-boundary correspondence. For $v/w < 1$, two bands of delocalized bulk states (low IPR, purple) are gapped. Within this gap lies a zero-energy state with a high IPR (yellow), confirming its localization at the chain's edges. This state vanishes for $v/w > 1$.

\subsection{Competition between Topology and Quasi-periodic Disorder}
\begin{figure}[h!]
    \centering
    \includegraphics[width=\textwidth]{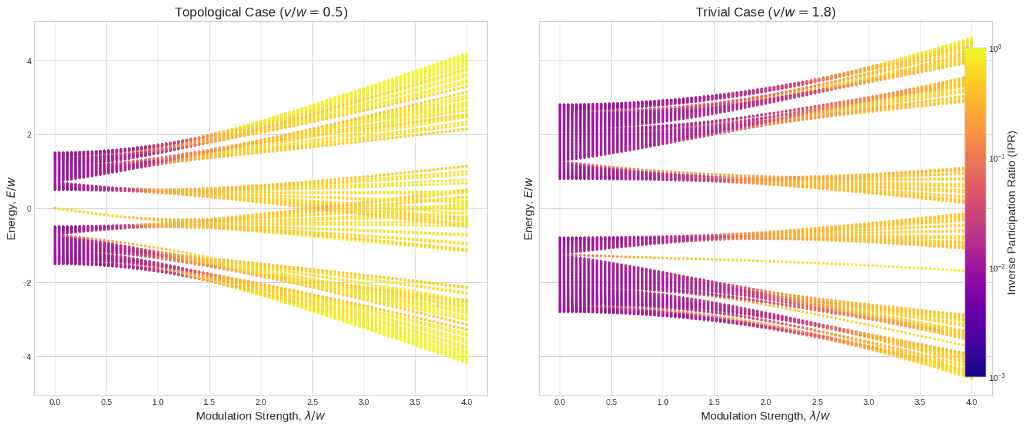}
    \caption{Energy spectrum versus AAH modulation strength $\lambda/w$, colored by IPR. (Left) Topological case. (Right) Trivial case. In both scenarios, a localization transition occurs around $\lambda/w=2$, where all states become localized.}
    \label{fig:ssh_aah}
\end{figure}

Next, we investigate the SSH-AAH model. Figure \ref{fig:ssh_aah} shows the energy spectrum versus the modulation strength $\lambda/w$ for a topological ($v/w=0.5$) and a trivial ($v/w=1.8$) chain.

As $\lambda$ increases, the spectrum evolves into a Hofstadter butterfly pattern. As $\lambda/w$ approaches 2, all states transition from delocalized (purple) to localized (yellow). This demonstrates that strong quasi-periodic disorder induces a localization transition that is independent of the underlying topology.

\subsection{Machine Learning-based Phase Classification}
\begin{figure}[h!]
    \centering
    \includegraphics[width=0.7\textwidth]{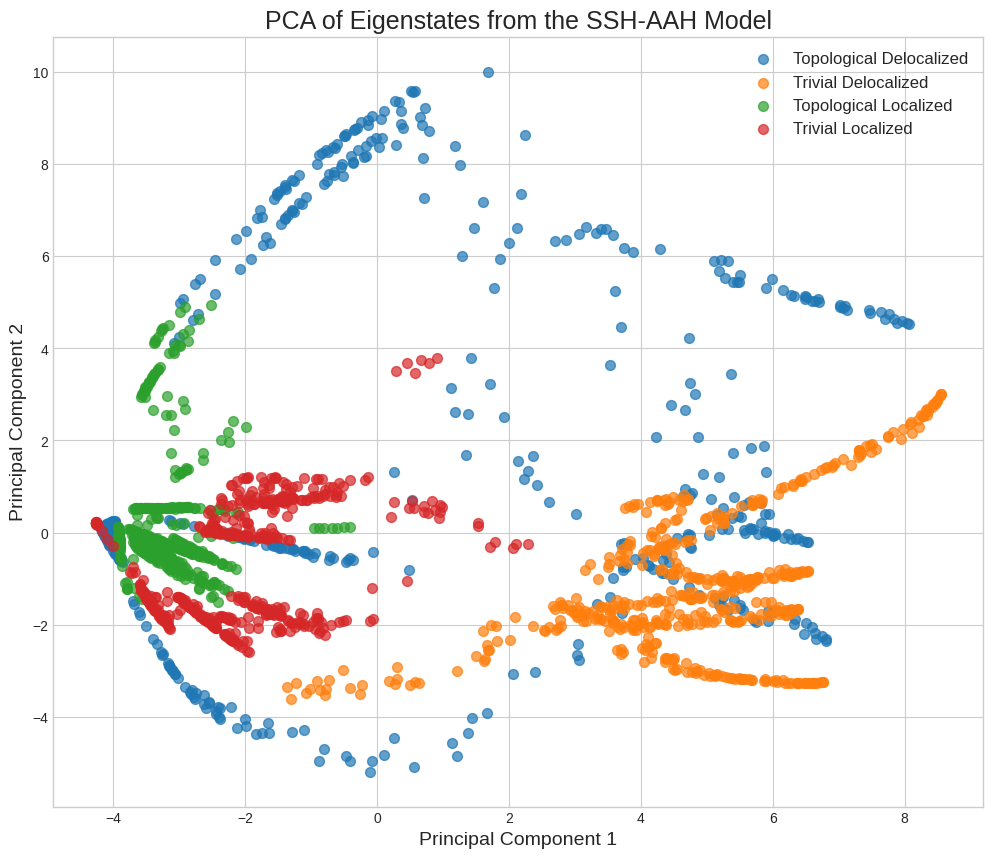}
    \caption{PCA of the eigenstates from the SSH-AAH model. PC1 (x-axis) clearly distinguishes localized from delocalized states. PC2 (y-axis) helps separate topological from trivial states in the delocalized regime.}
    \label{fig:pca}
\end{figure}

We employed PCA to classify the eigenstates from the four regimes of the SSH-AAH model (Fig.~\ref{fig:pca}).

The first principal component (PC1) provides a near-perfect separation between delocalized states (left) and localized states (right), indicating localization is the dominant feature. In the delocalized regime, PC2 separates the topological (blue) from trivial (green) states. The arc-like distribution of these points likely reflects the smooth variation of the underlying Bloch wavefunctions as a function of their crystal momentum. Most interestingly, in the localized regime, the topological (red) and trivial (orange) states are heavily intermixed, suggesting that strong localization masks the more subtle topological signatures.

\subsection{The Non-Hermitian Skin Effect}
\begin{figure}[h!]
    \centering
    \includegraphics[width=\textwidth]{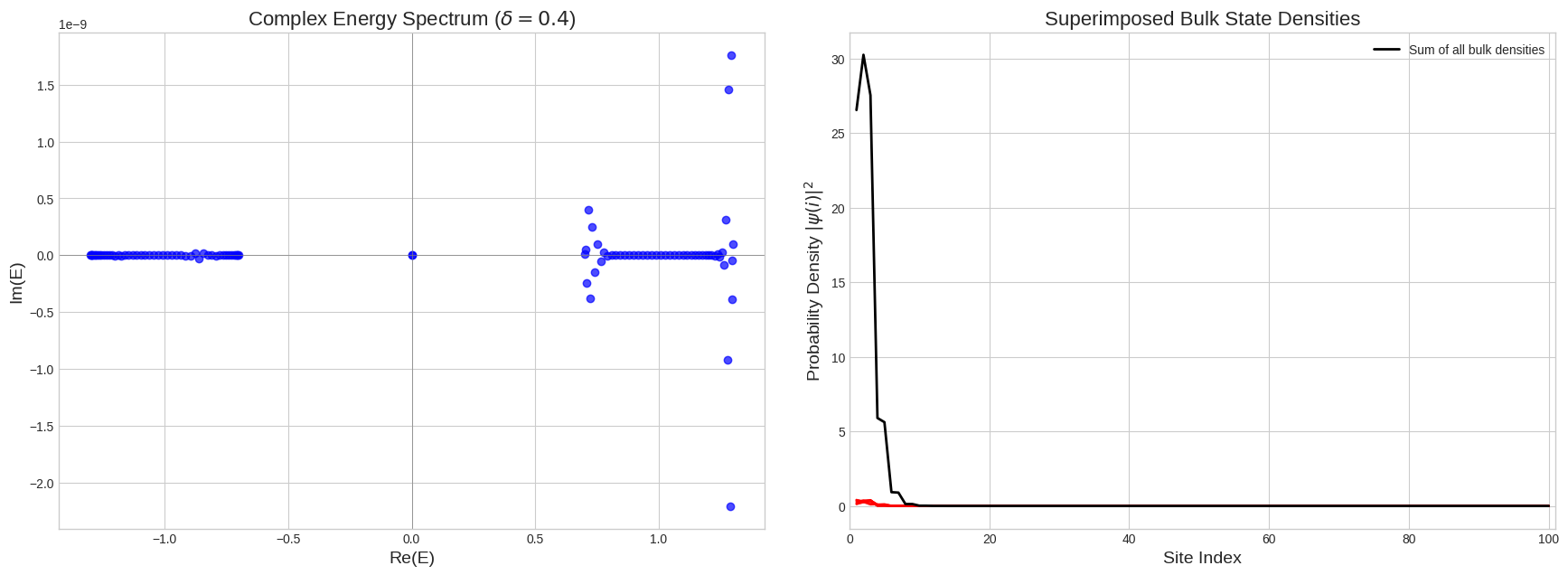}
    \caption{(Left) The complex energy spectrum of the non-Hermitian SSH chain. (Right) The superimposed probability densities of all bulk eigenstates, showing a massive accumulation at the left boundary. This is the non-Hermitian skin effect.}
    \label{fig:nh_skin}
\end{figure}

Moving beyond Hermitian systems, we introduced non-reciprocal hopping ($\delta=0.4$) into the chain. The results are in Fig.~\ref{fig:nh_skin}.

The left panel confirms the complex energy spectrum. The right panel displays the non-Hermitian skin effect: all bulk eigenstates are localized at the left edge. This accumulation occurs on the left due to our choice of a positive $\delta$; reversing its sign would cause the states to accumulate on the right edge. This phenomenon signals a fundamental breakdown of the conventional bulk-boundary correspondence.

\subsection{Floquet Engineered Topological Phases}
\begin{figure}[h!]
    \centering
    \includegraphics[width=0.7\textwidth]{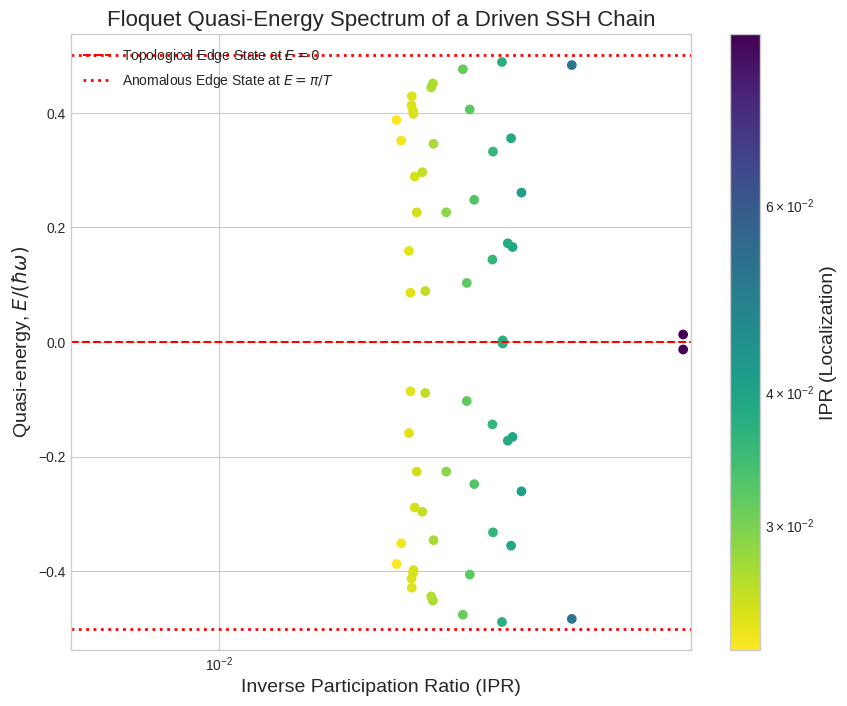}
    \caption{The quasi-energy spectrum of a driven trivial SSH chain, plotted against IPR. The drive induces localized states (high IPR) not only at $E=0$ but also at the edges of the Floquet zone, $E=\pm\omega/2$ (or $\pm\pi/T$). These are the anomalous Floquet edge states.}
    \label{fig:floquet}
\end{figure}

Finally, we applied a periodic drive to a trivial SSH chain. The resulting Floquet quasi-energy spectrum is plotted against IPR in Fig.~\ref{fig:floquet}.

The drive successfully induced a non-trivial topological phase. Figure \ref{fig:floquet} shows the emergence of localized states (high IPR). We find two states at quasi-energy $E=0$ and two additional localized states at $E \approx \pm\omega/2$. These latter states are the "anomalous edge states," a signature of a uniquely Floquet topological phase with no static counterpart.

\section{Conclusion}
\label{sec:conclusion}

In this work, we conducted a computational study of the SSH model under generalizations including quasi-periodic disorder, non-Hermiticity, and time-dependent driving. Our findings illustrate that moving beyond the ideal SSH model opens the door to a vast and complex landscape of quantum phenomena. We demonstrated that topological protection succumbs to a universal localization transition under strong disorder. Our machine learning analysis revealed that strong localization can mask underlying topological signatures. Furthermore, we showed that non-Hermiticity introduces the skin effect, a distinct localization mechanism, while Floquet engineering can dynamically create novel topological phases in trivial systems, evidenced by anomalous edge states. This work highlights the power of computational modeling as an indispensable tool for exploring these frontiers of condensed matter physics and provides a foundation for future studies into even more complex systems.


\end{document}